\documentclass[useAMS,fleqn,onecolumn]{mn2e}
\usepackage{amssymb}
\usepackage{amsmath}
\usepackage{graphicx}
\usepackage{times}

\newcommand{\calE}{{\mathcal{E}}}
\newcommand{\calN}{{\mathcal{N}}}
\newcommand{\calX}{{\mathcal{X}}}

\newcommand{\bfr}{{\boldsymbol{r}}}
\newcommand{\bfv}{{\boldsymbol{v}}}
\newcommand{\txA}{{\text{A}}}
\newcommand{\txQ}{{\text{Q}}}
\newcommand{\txR}{{\text{R}}}
\newcommand{\txa}{{\text{a}}}
\newcommand{\txd}{{\text{d}}}
\newcommand{\txp}{{\text{p}}}

\renewcommand{\leq}{\leqslant}

\DeclareMathOperator{\arctanh}{arctanh}
\DeclareMathOperator{\arcsinh}{arcsinh}
\DeclareMathOperator{\arccosh}{arccosh}

\title[Dynamical models for galaxies and bulges with a black hole]{%
A completely analytical family of dynamical models for spherical
galaxies and bulges with a central black hole}

\author[M. Baes \& H. Dejonghe]{Maarten Baes\thanks{Postdoctoral Fellow of
the Fund for Scientific Research, Flanders, Belgium
(F.W.O.-Vlaanderen)} and Herwig Dejonghe \\ Sterrenkundig Observatorium,
Universiteit Gent, Krijgslaan 281-S9, B-9000 Gent, Belgium}

\begin{document}

\maketitle

\begin{abstract}
We present a family of spherical models for elliptical galaxies and
bulges consisting of a stellar component and a central black hole. All
models in this family share the same stellar density profile, which
has a steep central cusp. The gravitational potential of each models
is a linear combination of the potential generated selfconsistently by
the stars and the potential of a central black hole. The relative
importance of these two contributions is a free parameter in the
models. Assuming an isotropic dynamical structure, almost all
kinematical properties of these models can be calculated
analytically. In particular, they form the first simple dynamical
models for galaxies with a central black hole where the distribution
function and differential energy distribution can be written
completely in terms of elementary functions only. We also present
various extensions of this family to models with anisotropic orbital
structures. Also for these dynamical models, the distribution function
and its moments can be expressed completely in terms of elementary
functions. 

This family is useful for a large range of applications, in particular
to generate initial conditions for N-body and hydrodynamical
simulations to model galactic nuclei with a central black hole.
\end{abstract}


\section{Introduction}

During the past few years, various numerical dynamical modelling
techniques have been developed to construct accurate dynamical models
for galaxies (Dejonghe~1989; Emsellem, Monet \& Bacon~1994; Rix et
al.~1997; van der Marel et al.~1998; Gerhard et al.~1998; Cretton et
al.~1999; Gebhardt et al.~2000a; Verolme \& de Zeeuw~2002). The
state-of-the-art techniques can handle various degrees of complexity
such as the capability to include non-trivial geometries and
higher-order kinematical information. Nevertheless, analytical models
remain interesting and important for various reasons. They can provide
a simple testbed where various physical processes can be investigated,
or where new modelling or data analysis techniques can be
explored. Analytical models are particularly useful to generate the
initial conditions for N-body, SPH or Monte Carlo simulations. Most
attention has been devoted to the construction of spherical
selfconsistent models, i.e.\ models in which the stars move in a
potential generated completely by the stars themselves. Famous
examples include the Plummer model (Plummer~1911; Dejonghe~1987), the
isochrone sphere~(H\'enon~1959, 1960) and the Jaffe sphere
(Jaffe~1983) and the Hernquist model (Hernquist~1990; Baes \&
Dejonghe~2002). Such models often serve as a template model for
elliptical galaxies, globular clusters or the bulges of spiral
galaxies.

Recent observational evidence indicates, however, that the central
regions of these stellar systems cannot always be faithfully
represented by selfconsistent models. The existence of supermassive
black holes in the nuclei of galaxies has been suspected for a long
time, as accretion onto massive compact objects was regarded as the
only reasonable explanation for the existence of active galaxies and
quasars (Salpeter~1964; So{\l}tan~1982). During the last decade,
quiescent supermassive black holes have been detected in the centre of
the Milky Way (Ghez et al.~2000; Sch\"odel et al.~2002) and in the
nuclei of virtually all external galaxies which are nearby enough to
resolve the sphere of influence of the black hole (e.g.~see list in
Tremaine et al.~2002). The masses of these black holes are roughly
between a million and a few billion solar masses and are tightly
coupled to structural parameters of the host galaxies (Kormendy \&
Richstone~1995; Ferrarese \& Merritt~2000; Gebhardt et al.~2000b;
Graham et al.~2001; McLure \& Dunlop~2002; Ferrarese~2002; Baes et
al.~2003; Marconi \& Hunt~2003). Recently, evidence for the presence
of intermediate mass black holes in the centre of globular clusters
has been reported (Gebhardt, Rich \& Ho~2002; Gerssen et al.~2002,
2003), although this evidence is still controversial (Baumgardt et
al.~2003a,b; McNamara, Harrison \& Anderson~2003). These observations
clearly indicate that there is a need for simple but realistic
analytical dynamical models with a central black hole.

The first requirement for the construction of models with a black hole
is a potential-density pair with a cuspy central density
profile. Tremaine et al.~(1994) demonstrate that spherical isotropic
systems need a central density profile that diverges at least as
$r^{-1/2}$ to be able to support a central black hole. The need for
such cuspy models also follows from observations: HST photometry has
shown that ellipticals and the bulges of spiral galaxies generally
have central density cusps that diverge as $r^{-\gamma}$ at small
radii with $0\leq\gamma\leq2.5$ (Lauer et al.~1995; Gebhardt et
al.~1996; Ravindranath et al.~2002; Seigar et al.~2002). Also for the
Milky Way there is evidence for a central density cusp in the
innermost regions (Genzel et al.~2003). A very useful one-parameter
family of spherical models with this characteristic, which we refer to
as the $\gamma$-models, was introduced independently by Dehnen~(1993)
and Tremaine et al.~(1994). This family has a simple analytical
density profile which diverges as $r^{-\gamma}$ in the central regions
$(0\leq\gamma<3)$, and includes the Hernquist and Jaffe models as
special cases. Zhao~(1996) generalized this family further to a very
general three-parameter family of models, the so-called
$(\alpha,\beta,\gamma)$-models. For many of these models, interesting
dynamical properties such as the intrinsic and projected velocity
dispersions, the distribution function and the differential energy
distribution can be calculated analytically if one assumes an
isotropic selfconsistent dynamical structure.

The second step in the construction of dynamical models with a cuspy
core is to add to the potential of the selfconsistent model an extra
contribution from the black hole, and re-calculate the dynamical
properties with this new potential. The calculation of the (intrinsic
and/or projected) velocity dispersions in the presence of a black hole
is not very difficult, and can usually be performed analytically for
those models where the dispersions can be calculated analytically in
the selfconsistent case. For example, for the sets of
potential-density pairs considered by Tremaine et al.~(1994) and
Zhao~(1996), the addition of a black hole was not a problem for the
calculation of the velocity dispersion profile. The reason is that the
intrinsic and projected velocity dispersions are just linear functions
of the potential, and therefore linear functions of the black hole
mass. Many other interesting kinematical properties, in particular the
phase space distribution function, depend on the potential in a
strongly non-linear way, however. The construction of dynamical models
in which these more complicated kinematical quantities can be
expressed analytically in the presence of a black hole proves to be
more difficult. Apart from the asymptotic behaviour, these
characteristics usually need to be calculated numerically (Tremaine et
al.~1994; Dejonghe et al.~1999).

The work by Ciotti~(1996) provides a first attempt to construct
completely analytical dynamical models for galaxies with a central
black hole. In an effort to construct realistic dynamical models for
elliptical galaxies embedded in massive cuspy dark matter haloes, he
considers a set of two-component Hernquist models. They consist of a
stellar component and a halo component, both of which are modelled as
a Hernquist profile, but with different masses and core radii. The
interesting aspect of his work for our goal is that the masses and
core radii of each component can be taken arbitrarily. Setting the
core radius of the halo component to zero, his two-component model
degenerates into a Hernquist model with a central black hole. Ciotti
demonstrates that the distribution function and differential energy
distribution of Hernquist models with a central black hole can be
expressed analytically, albeit as rather cumbersome expressions. For
example, the distribution function can be written as the derivative of
a complicated function involving incomplete elliptic integrals and
Jacobian functions. Nevertheless, his work presents the first model
for galaxies with a central black hole where most of the kinematical
properties can be calculated analytically.

In this paper, we present a detailed kinematical analysis of a family
of spherical models for elliptical galaxies and bulges consisting of a
stellar component and a central black hole. In Section~2 we define our
family of models. All models in this one-parameter family share the
same stellar density profile, which corresponds to one of
$\gamma$-models introduced by Dehnen~(1993) and Tremaine et
al.~(1994). The potential of the models is a linear combination of the
potential generated by the stars and the potential of a central black
hole. The importance of both components to the total potential can be
set arbitrarily by varying the parameter $\mu$, which represents the
mass of the central black hole relative to the total mass. The reason
why we chose this particular family of models is that most of the
interesting kinematical properties can be expressed completely
analytically for all values of $\mu$. In Section~3 we describe some
intrinsic kinematical properties of this family of models, such as the
distribution function, the differential energy distribution and the
moments of the distribution function. In particular, we give closed
analytical expressions for these quantities, completely in terms of
elementary functions. In Section~4 we describe some observed
kinematical properties as they are projected on the plane of the
sky. Most of these quantities can be expressed analytically involving
simple incomplete elliptic integrals. In Section~5 we present a number
of generalizations to models with a anisotropic orbital structure. In
particular, we discuss models with a constant anisotropy profile,
models with a distribution function of the Osipkov-Merritt type and
models with a completely radial orbital structure. Also for these
models, the distribution function and its moments can be calculated
completely analytically for all values of the central black hole
mass. Finally, a discussion is given in Section~5.

\section{Definition of the models}

The starting point for our family of models is the stellar luminosity
density profile
\begin{equation}
	\rho(r)
	=
        \frac{L_\star}{8\pi}
	\left(\frac{r}{c}\right)^{-5/2}
	\left(1+\frac{r}{c}\right)^{-3/2},
\end{equation}	
where $L_\star$ is the total stellar luminosity of the system and $c$
is a scale length. The (positive) potential generated by this stellar
distribution is easily found through Poisson's equation,
\begin{equation}
	\Psi_\star(r)
	=
	-\Phi_\star(r)
	=
	\frac{2GM_\star}{c}
    	\left(\sqrt{\frac{r+c}{r}}-1\right),
\end{equation}
where $M_\star$ is the total stellar mass. The isotropic
selfconsistent model with this potential-density pair is a particular
case from the family of $\gamma$-models considered by Dehnen~(1993)
and Tremaine et al.~(1994), corresponding to $\gamma=\tfrac{5}{2}$ or
$\eta=\tfrac{1}{2}$. A number of kinematical properties for this
selfconsistent model, such as the velocity dispersion and distribution
function has been derived in these papers. Our aim is to consider
systems consisting of two components: a stellar component and a
central supermassive black hole. In order to find the total potential
of such a two-component system, we need to add to the stellar
potential an extra contribution from the black hole. We model the
central black hole as a point potential with mass $M_\bullet$, such
that we obtain
\begin{equation}
    	\Psi(r)
    	=
	\frac{2GM_\star}{c}
    	\left(\sqrt{\frac{r+c}{r}}-1\right)
    	+
    	\frac{GM_\bullet}{r}.
\end{equation}
In the remainder of this paper, we will work in dimensionless units:
we set the gravitational constant $G$, the scalelength $c$, the
luminosity $L_\star$ and the {\em total} mass $M_\star+M_\bullet$ of
the model equal to unity. We introduce the parameter $\mu$ as the
fraction of the black hole mass to the total mass of the galaxy. With
these conventions, we can write the potential-density pair of our
model as
\begin{gather}
        \rho(r)
        =
        \frac{1}{8\pi}\,
    	\frac{1}{r^{5/2}\,(1+r)^{3/2}},
\label{density}
	\\
    	\Psi(r)
	=
    	2\left(1-\mu\right)\left(\sqrt{\frac{1+r}{r}}-1\right)
    	+
    	\frac{\mu}{r}.
\label{potential}
\end{gather}
In this paper, we consider isotropic kinematical models implied by the
potential-density pair formed by equations~(\ref{density}) and
(\ref{potential}). This potential-density pair defines a one-parameter
family of isotropic dynamical models, where $\mu$ is a free parameter
that can assume any value between 0 and 1. The models corresponding to
these values of $\mu$ have a particular meaning: the former
corresponds to a selfconsistent model without a central black hole,
whereas the latter represents a galaxy with a central black hole and a
negligible stellar mass-to-light ratio.

As an important remark, the reader should note that the convention we
use is different from the convention used in e.g.~Tremaine et
al.~(1994) and Zhao~(1996). In these papers, $\mu$ denotes the black
hole mass relative to the {\em stellar} mass, and the normalization is
such that the stellar mass is set to unity. We prefer to set the {\em
total} mass of the galaxy equal to unity however, because (1)~all
models then have the same behaviour at large radii and (2)~this allows
us to study the entire range of models from selfconsistent models
without black hole to models completely dominated by the black hole
potential.

\section{Intrinsic properties of the models}

\subsection{Basic properties}

\begin{figure*}
\centering
\includegraphics[clip,bb=104 49 290 685,angle=-90,width=\textwidth]{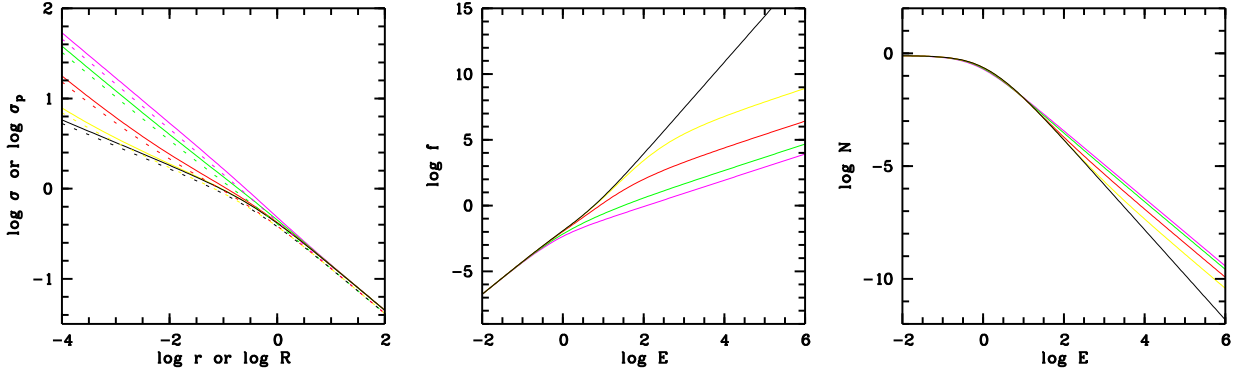}
\caption{
Some intrinsic and projected kinematical properties of the model for
various values of the parameter $\mu$. The first panel shows the
intrinsic velocity dispersion profile $\sigma(r)$ [solid lines] and
the projected velocity dispersion profile $\sigma_\txp(R)$ [dotted
lines], the second panel shows the distribution function $f(\calE)$
and in the right panel the differential energy distribution
$\calN(\calE)$ is plotted. The colour code is the same for all panels:
$\mu=0$ (black), $\mu=0.01$ (yellow), $\mu=0.1$ (red), $\mu=0.5$
(green) and $\mu=1$ (magenta).}
\label{all.eps}
\end{figure*}

The total mass density, circular velocity curve and the cumulative
mass function are readily obtained
\begin{gather}
	\rho_{\text{tot}}(r)
	=
        \frac{1-\mu}{8\pi}\,
    	\frac{1}{r^{5/2}\,(1+r)^{3/2}}
	+
	\mu\,\delta(\bfr),
	\\
    	v_{\text{c}}^2(r)
    	=
    	\frac{1-\mu}{\sqrt{r\,(1+r)}}+\frac{\mu}{r},
    	\\
    	M(r)
    	=
    	\left(1-\mu\right)\sqrt{\frac{r}{1+r}}+\mu.
\end{gather}
The isotropic velocity dispersion
$\sigma=\sigma_r=\sigma_\theta=\sigma_\phi$ can be obtained from the
solution of the Jeans equation,
\begin{equation}
    	\sigma^2(r)
    	=
    	\frac{1}{\rho(r)}
    	\int_r^\infty
    	\frac{M(r)\,\rho(r)\,\txd r}{r^2}.
\label{jeans}
\end{equation}
The cumulative mass function, which depends linearly on the black hole
mass, is the only $\mu$-dependent quantity in this
expression. Consequently we can write the velocity dispersion as the
sum of separate contributions from the stellar mass and the black
hole. The result is
\begin{align}
	\sigma^2(r) 
	&= 
	\frac{1-\mu}{3}\,	
	\sqrt{\frac{1+r}{r}}
	\left[
	1-2r+6r^2+12r^3
	-12r^3(1+r)\ln\left(\frac{1+r}{r}\right)
	\right]
	\nonumber \\
	&\qquad\qquad\qquad
	+\frac{2\mu}{35}
	\left(\frac{1+r}{r}\right)
    	\left(
	5-8r+16r^2-64r^3-128r^4
    	+128r^{7/2}\sqrt{1+r}
	\right).
\label{sigma}
\end{align}
At large radii, the velocity dispersion goes to zero as
\begin{equation}
    	\sigma^2(r)
    	\approx
    	\frac{1}{5r}
	\left(
	1-\frac{2-5\mu}{12r}
	+\cdots
	\right).
\end{equation}
At small radii, the velocity dispersion diverges as
\begin{equation}
    	\sigma^2(r)
    	\approx
    	\frac{1-\mu}{3\sqrt{r}}
	\left(
	1-\frac{3r}{5}+\cdots
	\right)
	+
    	\frac{2\mu}{7r}
	\left(
	1-\frac{3r}{2}+\cdots
	\right).
\end{equation}
In the left panel of figure~{\ref{all.eps}} we plot the intrinsic
velocity dispersion profile for various black hole masses. Without a
central black hole, the velocity dispersion (and the circular
velocity) have a weak $r^{1/4}$ divergence at small radii. When a
black hole is present, these divergences become stronger and both
quantities diverge as $r^{-1/2}$.

\subsection{The distribution function}
\label{df.sec}

The ultimate goal of the dynamical modeller is to obtain an expression
for the phase space distribution function $f(\bfr,\bfv)$, which
contains all kinematical information on a system. For non-rotating
isotropic spherically symmetric systems, the phase space distribution
function depends on the position and velocity vectors only through the
binding energy $\calE = \psi(r)-\tfrac{1}{2}\bfv^2$, which is a
integral of motion. The distribution function $f(\bfr,\bfv)=f(\calE)$
is thus a one-dimensional rather than a six-dimensional function. The
key to calculate the distribution function corresponding to a given
potential-density pair is the augmented density $\tilde\rho(\Psi)$,
i.e.\ the density written in terms of the potential. With this
function, we can calculate the distribution function through the
Eddington~(1916) relation,
\begin{equation}
        f(\calE)
        =
        \frac{1}{\sqrt{8}\pi^2}
        \left[
        \int_0^\calE \frac{\txd^2\tilde\rho}{\txd\Psi^2}\,
        \frac{\txd\Psi}{\sqrt{\calE-\Psi}}
        +
        \frac{1}{\sqrt{\calE}}
        \left(\frac{\txd\tilde\rho}{\txd\Psi}\right)_{\Psi=0}
        \right].
\label{eddington}
\end{equation}
The second term in this expression vanishes for the family of models
we consider in this paper, both with and without black hole, because
$\tilde\rho(\Psi)\propto \Psi^4$ at large radii. To find an expression
for the augmented density, we need to invert the potential to a
relation $r(\Psi)$ and determine $\tilde\rho(\Psi) = \rho(r(\Psi))$.

We first consider the special case when the galaxy contains no central
black hole ($\mu=0$). The augmented density is readily found,
\begin{equation}
    	\tilde\rho(\Psi) 
 	= 
	\frac{1}{256\pi}\,
    	\frac{\Psi^4\,(4+\Psi)^4}{(2+\Psi)^3},
\end{equation}
and the resulting distribution function is (Tremaine et al.~1994)
\begin{multline}
    	f(\calE)
    	=
    	\frac{\sqrt{2}}{896\pi^3}
    	\left[
    	\frac{42\,(3-32\calE-8\calE^2)}{(2+\calE)^{9/2}}
	\arcsinh\sqrt{\frac{\calE}{2}}
    	-
	\frac{(63-693\calE-5670\calE^2-7410\calE^3-4488\calE^4
    	-1448\calE^5-240\calE^6-16\calE^7)\sqrt{\calE}}
	{(2+\calE)^4}
    	\right].
\label{fnoBH}
\end{multline}
The behaviour at small binding energies is
\begin{equation}
    	f(\calE)
    	\approx
    	\frac{2\sqrt{2}}{5\pi^3}\,\calE^{5/2}
    	\left(1-\frac{5\calE}{7}+\cdots\right),
\label{fnoBHsmallE}
\end{equation}
whereas at large binding energies we obtain
\begin{equation}
    	f(\calE)
    	\approx
    	\frac{\sqrt{2}}{56\pi^3}\,\calE^{7/2}
    	\left(1+\frac{7}{\calE}+\cdots\right).
\label{fnoBHlargeE}
\end{equation}
Another special case is the other side of the spectrum ($\mu=1$),
which corresponds to a galaxy where the black hole completely
dominates the potential. In this case the augmented density and
distribution function read
\begin{gather}
    	\tilde\rho(\Psi) 
 	= 
	\frac{1}{8\pi}\,
    	\frac{\Psi^4}{(1+\Psi)^{3/2}},
	\\
    	f(\calE)
    	=
    	\frac{\sqrt{2}}{128\pi^3}
    	\left[
    	\frac{(9+9\calE+31\calE^2+15\calE^3)\,\sqrt{\calE}}{(1+\calE)^3}
	-
	3\left(3-5\calE\right)\arctan\sqrt{\calE}
	\right].
\label{fdomBH}
\end{gather}
The behaviour at small binding energies is
\begin{equation}
    	f(\calE)
    	\approx
    	\frac{2\sqrt{2}}{5\pi^3}\,\calE^{5/2}
    	\left(1-\frac{15\calE}{7}+\cdots\right),
\label{fdomBHsmallE}
\end{equation}
whereas at large binding energies we obtain
\begin{equation}
    	f(\calE)
    	\approx
    	\frac{15\sqrt{2}}{256\pi^2}\,\calE
    	\left(1-\frac{3}{5\calE}+\cdots\right).
\label{fdomBHlargeE}
\end{equation}
For the general case $0<\mu<1$, i.e.\ when both the stars and the
central black hole contribute to the potential of the galaxy, the
calculation of the distribution function is more tedious. In order to
find a convenient expression for the augmented density, we write
expression~(\ref{density}) as
\begin{equation}
	\rho(a)
	=
    	\frac{1}{8\pi}\,
    	\frac{(a^2-1)^4}{a^3}
	\qquad
	\text{with}
	\qquad
	a 
	\equiv 
	a(r)
	=
	\sqrt{\frac{1+r}{r}}.
\label{rhoa}
\end{equation}
A relation $a(\Psi)$ can be found by inverting the expression
\begin{equation}
	\Psi(a)
	=
	2\left(1-\mu\right)\left(a-1\right)
	+
	\mu\left(a^2-1\right),
\label{Psia}
\end{equation}
which yields
\begin{equation}
	a(\Psi)
	=
	\frac{\mu-1+\sqrt{1+\mu\Psi}}{\mu}. 
\end{equation}
To calculate a closed expression for the distribution function, we can
now take different approaches. The first approach is a direct
application of the Eddington formula~(\ref{eddington}). When we
substitute the explicit expression for $a(\Psi)$ into the
formula~(\ref{rhoa}), we find after some algebra that the second
derivative of the augmented density can be written in the form
\begin{equation}
    	\frac{\txd^2\tilde\rho}{\txd\Psi^2}(\Psi)
    	=
    	\frac{1}{32\pi\,\mu^3\,(2-\mu+\Psi)^5}
    	\left[
    	P_1(\Psi,\mu)
    	+
    	\frac{P_2(\Psi,\mu)}
    	{(1+\mu\Psi)^{3/2}}
    	\right],
\end{equation}
where $P_1$ and $P_2$ are polynomials with integer coefficients in
$\Psi$ and $\mu$. This form is suitable for analytical integration in
the Eddington relation: the two parts of the integrand are basically
the combination of a rational function of $\Psi$ and the square root
of first and second order polynomials in $\Psi$. Such a function can
be integrated analytically and the resulting integral can be written
in terms of elementary functions only. The distribution function can
also be calculated in a more elegant way by transforming the Eddington
relation~(\ref{eddington}) to an integration with respect to $a$. We
obtain
\begin{equation}
        f(\calE)
        =
        \frac{1}{\sqrt{8}\pi^2}
        \int_1^{a_\calE}
	\frac{\txd}{\txd a}
	\left[
	\left.
	\frac{\txd\tilde\rho}{\txd a}
	\right/
	\frac{\txd\Psi}{\txd a}
	\right]
        \frac{\txd a}{\sqrt{\calE-2(1-\mu)(a-1)-\mu(a^2-1)}}
\label{eddingtona}
\end{equation}
with
\begin{equation}
	a_\calE 
	= 
	\frac{\mu-1+\sqrt{1+\mu\calE}}{\mu}.
\label{acalE}
\end{equation}
With the expressions~(\ref{rhoa}) and (\ref{Psia}) the first factor in
the integrand becomes
\begin{align}
	\frac{\txd}{\txd a}
	\left[
	\left.
	\frac{\txd\tilde\rho}{\txd a}
	\right/
	\frac{\txd\Psi}{\txd a}
	\right]
	&=
	\frac{1}{16\pi}\,
	\frac{
	(a^2-1)^2\,
	[12(1-\mu)+15\mu a+16(1-\mu)a^2+18\mu a^3+20(1-\mu)a^4+15\mu a^5]
	}{
	a^5\,(1-\mu+\mu a)^2
	}.
\end{align}		
This function can be decomposed in partial fractions, and each of the
resulting integrals can be calculated analytically in terms of
elementary functions. Using either of both approaches, we obtain that
the resulting distribution function can be written explicitly as
\begin{multline}
    	f(\calE)
    	=
    	\frac{\sqrt{2}}{128\pi^3}
	\left[
    	-
	\frac{\calX(\calE,\mu)\,\sqrt{\calE}}
    	{\mu^3\,(2-\mu+\calE)^4\,(1+\mu\calE)}
	+
    	\frac{3\,(25-40\mu+12\mu^2+5\mu\calE)}{\mu^{7/2}}
    	\arctan\sqrt{\mu\calE}
    	\right.
	\\
	+
    	\left.
	\frac{3\,(1-\mu)\,(3-8\mu+12\mu^2-32\calE-31\mu\calE-8\calE^2)}
    	{(2-\mu+\calE)^{9/2}}
	\arctanh\left(\frac{\sqrt{(2-\mu+\calE)\calE}}{1+\calE}\right)
    	\right],
\label{fBH}
\end{multline}
where $\calX(\calE,\mu)$ is a polynomial with integer coefficients
\begin{align}
    	\calX(\calE,\mu)
    	&=
    	3(400-1440\mu+2072\mu^2-1541\mu^3+622\mu^4-116\mu^5)
    	+(2400-6400\mu-5352\mu^2-503\mu^3-1658\mu^4+1002\mu^5+220\mu^6)\,\calE
    	\nonumber\\
    	&
    	+(1800-2600\mu-1486\mu^2+4030\mu^3-2175\mu^4+373\mu^5)\,\calE^2
    	+2\,(300+150\mu-1356\mu^2+1116\mu^3-253\mu^4)\,\calE^3
    	\nonumber\\
    	&
    	+(75+400\mu-864\mu^2+328\mu^3)\,\calE^4
    	+5\mu\,(13-16\mu)\,\calE^5.
\end{align}
The first two terms in the distribution function~(\ref{fBH}) both
diverge as $\mu^{-3}$ when $\mu$ approaches zero, but the diverging
terms of course cancel out, such that the distribution function
reduces to the expression~(\ref{fnoBH}) in the limit
$\mu\rightarrow0$, as required. On the other hand, it is
straightforward to check that the distribution function~(\ref{fBH})
reduces to the expression~(\ref{fdomBH}) in the limit
$\mu\rightarrow1$. At small binding energies, the distribution
function~(\ref{fBH}) behaves as
\begin{equation}
    f(\calE)
    \approx
    \frac{2\sqrt{2}}{5\pi^3}\,\calE^{5/2}
    \left[1-\frac{5\,(1+2\mu)\calE}{7}+\cdots\right],
\label{fBHsmallE}
\end{equation}
and at large binding energies we have the expansion
\begin{equation}
   	 f(\calE)
    	\approx
    	\frac{15\sqrt{2}}{256\pi^2}\,
	\frac{\calE}{\mu^{5/2}}\,
    	\left[
	1
	-\frac{32(1-\mu)}{3\pi\sqrt{\mu}\,\sqrt{\calE}}
	+\frac{25-40\mu+12\mu^2}{5\mu\,\calE}
	+\cdots\right].
\label{fBHlargeE}
\end{equation}
In the second panel of figure~{\ref{all.eps}} we plot the distribution
function for different values of the black hole mass. As the central
potential well is infinitely deep, orbits with all binding energies
are allowed. Both without and with a central black hole, the
distribution function is a monotonically rising function, which
guarantees the stability of the model against radial and non-radial
perturbations. At small binding energies the influence of a black hole
is negligible. This is also visible in the asymptotic
expansion~(\ref{fBHsmallE}), where the leading term is independent of
the black hole mass fraction $\mu$. The same leading terms is then
obviously found in the asymptotic expansions~(\ref{fnoBHsmallE})
and~(\ref{fdomBHsmallE}). At large binding energies, i.e.\ in the
central regions of the system, the influence of the black hole is
important. The distribution function becomes less steep in the large
binding energy limit when a black hole is present: its slope changes
from $\tfrac{7}{2}$ to $1$.  The larger the black hole mass, the
smaller the values of the distribution function.

\subsection{The differential energy distribution}

The distribution function $f(\bfr,\bfv) = f(\calE)$ contains all
available kinematical information on the system, but is rather hard to
interpret. In particular, $f(\calE)$ does {\em not} represent the
number of stars per unit binding energy. The quantity that describes
this useful characteristic is the differential energy distribution
$\calN(\calE)$. For an isotropic spherical system, the differential
energy distribution can be written as $\calN(\calE) =
f(\calE)\,g(\calE)$, where $g(\calE)$ is called the density of states
and represents the phase space volume per unit binding energy (Binney
\& Tremaine~1987). This function can be calculated through the
equation
\begin{equation}
	g(\calE)
	=
	16\sqrt{2}\pi^2 
	\int_{\calE}^\infty
	\left|r^2\frac{\txd r}{\txd\Psi}\right|
	\sqrt{\Psi-\calE}\,
	\txd\Psi.
\label{defdos}
\end{equation}
For the selfconsistent model without black hole, the density of states
can immediately be calculated
\begin{equation}
	g(\calE)
	=
	\sqrt{2}\pi^3
	\left[
	\frac{1-\calE+\calE^2}{\calE^{5/2}}
	-
	\frac{21+9\calE+\calE^2}{(\calE+4)^{5/2}}
	\right],
\label{dosnoBH}
\end{equation}
and the differential energy distribution is found by multiplying this
expression with the distribution function~(\ref{fnoBH}). At small
binding energies, the differential energy distribution asymptotically
behaves as
\begin{equation}
	\calN(\calE)
	\approx
	\frac{4}{5}
	\left(
	1-\frac{12\calE}{7}+\cdots
	\right),
\label{dednoBHsmallE}
\end{equation}
and at large binding energies we have the expansion
\begin{equation}
	\calN(\calE)
	\approx
	\frac{3}{2\calE^2}
	\left(1-\frac{4}{\calE}+\cdots\right).
\label{dednoBHlargeE}
\end{equation}
If the potential of the galaxy is completely dominated by the black
hole, we obtain the well-known and very simple expression for the
density of states function,
\begin{equation}
	g(\calE)
	=
	\frac{\sqrt{2}\,\pi^3}{\calE^{5/2}}.
\label{dosdomBH}
\end{equation}
Combining this expression with equation~(\ref{fdomBH}) for the
distribution function, we find that the differential energy
distribution asymptotically behaves as
\begin{equation}
	\calN(\calE)
	\approx
	\frac{4}{5}
	\left(
	1-\frac{15\calE}{7}+\cdots
	\right),
\label{deddomBHsmallE}
\end{equation}
in the small binding energy limit, whereas at large binding energies
we have the asymptotic expansion
\begin{equation}
	\calN(\calE)
	\approx
	\frac{15\pi}{128\calE^{3/2}}
	\left(1-\frac{3}{5\calE}+\cdots\right).
\label{deddomBHlargeE}
\end{equation}
In the general case $0<\mu<1$, the calculation of the density of
states is less straightforward. Similarly as for the distribution
function, we have calculated the function $g(\calE)$ in two ways. The
first approach uses a brute force calculation of the
definition~(\ref{defdos}). A more subtle approach consists of
transforming the integral in this expression to an integration with
respect to $a$. We obtain
\begin{equation}
	g(\calE)
	=
	32\sqrt{2}\pi^2 
	\int_{a_\calE}^\infty
	\frac{a}{(a^2-1)^4}\,
	\sqrt{2(1-\mu)(a-1)+\mu(a^2-1)-\calE}\,
	\txd a,
\label{defdosa}
\end{equation}
where $a_\calE$ as in equation~(\ref{acalE}). This integral can be
calculated analytically and results in
\begin{align}
	g(\calE)
	&=
	\sqrt{2}\pi^2
	\left\{
	\frac{
	2(1-\mu)\left[
	24(1-\mu)
	-4(3-10\mu+4\mu^2)\calE
	-2(9-11\mu)\calE^2
	-3\calE^3
	\right]\sqrt{\mu}
	}{
	3\calE^2(4-4\mu+\calE)^2
	}
	\right.
	\nonumber \\
	&\qquad\qquad\qquad
	-
	\frac{1-(1-\mu)\calE+(1-\mu)\calE^2}{\calE^{5/2}}
	\left[\pi-\arctan\sqrt{\mu\calE}\right]
	\nonumber \\
	&\qquad\qquad\qquad
	\left.
	+
	\frac{
	(21-70\mu+80\mu^2-32\mu^3)
	+(9-19\mu+10\mu^2)\calE
	+(1-\mu)\calE^2
	}{
	(4-4\mu+\calE)^{5/2}
	}
	\left[
	\arctan\left(\frac{2\mu-1}{\sqrt{\mu(4-4\mu+\calE)}}\right)
	-
	\frac{\pi}{2}
	\right]
	\right\}.
\label{dosBH}
\end{align}
In the limits $\mu\rightarrow0$ and $\mu\rightarrow1$, this function
reduces to the expressions~(\ref{dosnoBH}) and~(\ref{dosdomBH})
respectively. The differential energy distribution $\calN(\calE)$ of
the model we consider can thus be written completely in terms of
elementary functions for all values of $\mu$. At small binding
energies, the differential energy distribution behaves as
\begin{equation}
	\calN(\calE)
	\approx
	\frac{4}{5}
	\left[
	1
	-
	\frac{3(4+\mu)\,\calE}{7}
	+
	\cdots
	\right],
\end{equation}
and at large binding energies we find the expansion
\begin{equation}
	\calN(\calE)
	\approx
	\frac{15\pi\,\sqrt{\mu}}{128\,\calE^{3/2}}
	\left[
	1 
	+ 
	\frac{32(1-\mu)}{5\pi\,\sqrt{\mu}}\,\frac{1}{\sqrt{\calE}}
	+
	\left(
	\frac{100-215\mu+112\mu^2}{5\mu}
	-
	\frac{8192(1-\mu)^2}{45\pi^2\,\mu}
	\right)
	\frac{1}{\calE}
	+\cdots
	\right].
\end{equation}
In the right-hand panel of figure~{\ref{all.eps}} we plot the
differential energy distribution for models with various black hole
masses. Without a black hole, the differential energy distribution is
a monotonically decreasing function that has a finite value at
$\calE=0$ and that decreases smoothly to zero as $\calE^{-2}$ at large
binding energies. In spite of the strong divergence of the
distribution function at large binding energies, the system thus has
no stars at rest in the centre of the galaxy. Both of these properties
(a finite value at $\calE=0$ and convergence to zero at large $\calE$)
are common properties for all selfconsistent models in the family of
$\gamma$-models (Dehnen~1993). Adding a supermassive black hole does
not change the behaviour of the differential energy distribution very
drastically. The differential energy distribution still reaches the
same finite value $\tfrac{4}{5}$ at $\calE=0$, and then smoothly
decreases for increasing binding energies. The slope of the
differential energy distribution is a weakly decreasing function of
$\mu$. At large binding energies, the differential energy distribution
still converges to zero, but the slope of the convergence is weakened
from $-2$ to $-\tfrac{3}{2}$. For increasing black hole mass, the
differential energy distribution assumes smaller values at small
binding energies and converges less rapidly to zero at large binding
energies. Models with increasing black hole masses hence contain stars
which are in the mean increasingly strongly bound to the galaxy.

\subsection{The moments of the distribution function}

It is useful to consider the moments of the distribution function with
respect to the velocities. In a general non-rotating spherical system,
the moments of the distribution function are defined as
\begin{equation}
	\mu_{2i,2j,2k}(r)
	=
	\iiint f(r,\bfv) 
	v_r^{2i}\,
	v_\theta^{2j}\,
	v_\phi^{2k}\,\txd\bfv,
\end{equation}
where the integration covers the entire velocity space. For isotropic
spherical systems the distribution function is symmetric in the
velocities, and one defines the isotropic moments as
\begin{equation}
	\mu_{2n}(r)
	=
	4\pi\int_0^{\sqrt{2\Psi(r)}} f(r,v)\,v^{2n+2}\,\txd v.
\end{equation}
It is easily shown that the relation between the general moments and
the isotropic moments is
\begin{equation}
	\mu_{2i,2j,2k}(r)
	=
	\frac{1}{2\pi}\,
	\frac{\Gamma(i+\tfrac{1}{2})\,
	\Gamma(j+\tfrac{1}{2})\,
	\Gamma(k+\tfrac{1}{2})}
	{\Gamma(i+j+k+\tfrac{3}{2})}\,
	\mu_{2(i+j+k)}(r).
\end{equation}
Knowledge of the isotropic moments is thus sufficient to determine all
the moments. A direct integration of the definition is not the most
obvious way to calculate the moments. A more interesting way is to use
a formula that expresses the augmented moments, i.e.\ the moments
written as a function of the potential, as a function of the augmented
density (Dejonghe~1986),
\begin{equation}
	\tilde\mu_{2n}(\Psi)
	=
	\frac{(2n+1)!!}{(n-1)!!}
	\int_0^\Psi
	\left(\Psi-\Psi'\right)^{n-1}\,
	\tilde\rho\left(\Psi'\right)\,\txd\Psi'.
\end{equation}
For the model we consider, we can write this expression as
\begin{equation}
	\tilde\mu_{2n}(\Psi)
	\equiv
	\tilde\mu_{2n}(a)
	=
	\frac{1}{4\pi}\,
	\frac{(2n+1)!!}{(n-1)!!}
	\int_1^a
	\frac{\left(1-a'{}^2\right)^4\left(1-\mu+\mu a'\right)}{a'{}^3}\,
	\left[
	2\left(1-\mu\right)\left(a-a'\right)+\mu\left(a^2-a'{}^2\right)
	\right]^{n-1}\,
	\txd a'.
\end{equation}
For all integer values of $n$, the integrand of this integral is a
simple power series in $a'$ and can easily be integrated
analytically. The $2n$'th moment can be written as a polynomial in
$\mu$ of order $n$, where each coefficient is a sum of powers in $a$
and terms in $\ln a$. The true isotropic moments $\mu_{2n}(r)$ can
subsequently be found by substitution of $a=\sqrt{(1+r)/r}$ into the
result. For example, for the second order moment, one obtains
\begin{equation}
	\tilde\mu_2(a)
	=
	\frac{3}{\pi}
	\left[
	\left(1-\mu\right)
	\left(
	-\frac{1}{8a^2}
	-\frac{5}{12}
	-\ln a
	+\frac{3a^2}{4}
	-\frac{a^4}{4}
	+\frac{a^6}{24}
	\right)
	+
	\mu
	\left(
	-\frac{1}{a}
	+\frac{32}{35}
	-a
	+\frac{a^3}{2}
	-\frac{a^5}{5}
	+\frac{a^7}{28}
	\right)
	\right].
\end{equation}
If we substitute $a=\sqrt{(1+r)/r}$ and $\mu_2=3\rho\sigma^2$ into
this equation, we recover the expression~(\ref{sigma}).

\section{Observed properties of the model}

\subsection{The surface brightness}

The surface brightness profile can be found by projecting the
luminosity density profile on the plane of the sky,
\begin{equation}
    	I(R)
    	=
    	2\int_R^\infty
    	\frac{\rho(r)\,r\,\txd r}{\sqrt{r^2-R^2}}.
\label{Sigma}
\end{equation}
As all models in the family we investigate share the same stellar
density profile, they will obviously share the same surface brightness
profile. It can be written in terms of incomplete elliptic integrals
of the first and second kind,
\begin{equation}
    	I(R)
    	=
    	\frac{1}{\pi}
    	\left[
    	\frac{1}{2\,(1-R^2)}
    	+
    	\frac{R^2-\tfrac{1}{2}}{R^{3/2}\,(R-1)\,\sqrt{R+1}}\,
    	\mathbb{E}\left(\frac{\pi}{4},\sqrt{\frac{2}{R+1}}\right)
    	-
    	\frac{1}{\sqrt{R\,(R+1)}}\,
    	\mathbb{F}\left(\frac{\pi}{4},\sqrt{\frac{2}{R+1}}\right)
    	\right].
\label{Sigmahere}
\end{equation}
In the Appendix we give the relevant formulae to transform this
formula (and similar formulae which will appear in the remainder of
this discussion) to a more convenient form for $R<1$ and
$R\approx1$. At large projected radii, the surface brightness
decreases as
\begin{equation}
	I(R)
	\approx
	\frac{1}{16R^3}
	\left(
	1-\frac{4}{\pi\,R}+\cdots
	\right),
\end{equation}
and the behaviour at small projected radii is
\begin{equation}
    	I(R)
    	\approx
    	\sqrt{\frac{\pi}{2}}\,
    	\frac{1}{\Gamma^2}\,
    	R^{-3/2}
    	\left(
    	1
    	-
    	\frac{3\,\Gamma^4}{16\pi^2}R
    	+
    	\cdots
    	\right),
\end{equation}
where $\Gamma\equiv\Gamma(\tfrac{1}{4})\approx3.62561$. The cumulative
surface brightness profile can be found by integrating the surface
density over a circular surface on the plane of the sky,
\begin{equation}
    	S(R)
    	=
    	2\pi\int_0^R I(R')\,R'\,\txd R'.
\end{equation}
A more convenient expression can be found by substitution of the
definition~(\ref{Sigma}) into this expression and partial
integration,
\begin{equation}
    	S(R)
    	=
    	1-4\pi\int_R^\infty \rho(r)\,\sqrt{r^2-R^2}\,r\,\txd r.
\end{equation}
For the luminosity density profile we consider, this expression can
also be written in terms of incomplete elliptic integrals of the first
and second kind
\begin{equation}
    	S(R)
    	=
    	2\sqrt{R\,(R+1)}\,
    	\mathbb{E}\left(\frac{\pi}{4},\sqrt{\frac{2}{R+1}}\right)
    	-
    	\frac{2R^{3/2}}{\sqrt{R+1}}\,
    	\mathbb{F}\left(\frac{\pi}{4},\sqrt{\frac{2}{R+1}}\right).
\label{SR}
\end{equation}
The effective radius $R_{\text{eff}}$ is obtained by solving the
equation $S(R_{\text{eff}})=\tfrac{1}{2}$; a numerical evaluation
gives $R_{\text{eff}}=0.244955$.

\subsection{The projected velocity dispersion}

The projected velocity dispersion can be found by projecting the
intrinsic velocity dispersion profile on the plane of the sky,
\begin{equation}
    	\sigma_\txp^2(R)
    	=
    	\frac{2}{I(R)}
    	\int_R^\infty
    	\frac{\rho(r)\,\sigma^2(r)\,r\,\txd r}{\sqrt{r^2-R^2}},
\end{equation}
or after substitution of the Jeans solution~(\ref{jeans}) and partial
integration,
\begin{equation}
    	\sigma_\txp^2(R)
    	=
    	\frac{2}{I(R)}
    	\int_R^\infty
    	\frac{\rho(r)\,M(r)\,\sqrt{r^2-R^2}\,\txd r}{r^2}.
\label{nupsigmap2}
\end{equation}
Similarly as for the intrinsic dispersion, we can write the projected
velocity dispersion as the sum of separate contributions from the
stellar mass and the black hole. The contribution of the stellar mass
leads to an integral which can be written in terms of elementary
function, whereas the contribution from the black hole leads to a more
complicated integral which can be written in terms of the incomplete
elliptic integrals of the first and second kind,
\begin{align}
    	I(R)\sigma_\txp^2(R)
        &=
	(1-\mu)
	\left[
    	\frac{4R^2-1}{8R}
    	-
    	\frac{12R^2+1}{12\pi R^2}
    	+
    	\frac{3-4R^2}{4\pi}\,X(R)
	\right]
	\nonumber \\
	&\qquad
	+\frac{\mu}{105\pi}
    	\left[
    	\frac{5-96R^2}{R^2}
    	+
    	\frac{2\,(96R^2-17)\,\sqrt{R+1}}{R^{3/2}}\,
    	\mathbb{E}\left(\frac{\pi}{4},\sqrt{\frac{2}{R+1}}\right)
    	-
    	\frac{192R^4-82R^2-5}{R^{5/2}\sqrt{R+1}}\,
    	\mathbb{F}\left(\frac{\pi}{4},\sqrt{\frac{2}{R+1}}\right)
    	\right],
\label{sigp}
\end{align}
with $X(R)$ a continuous real function defined as
\begin{equation}
    	X(R)
    	=
    	\begin{cases}
    	\;
    	\dfrac{1}{\sqrt{1-R^2}}\arccosh\left(\dfrac{1}{R}\right)
    	&\qquad
    	\text{if $R<1$,}
    	\\[1em]
    	\;
    	\dfrac{1}{\sqrt{R^2-1}}\arccos\left(\dfrac{1}{R}\right)
   	&\qquad
    	\text{if $R>1$.}
    	\end{cases}
\end{equation}
At large projected radii, the projected velocity dispersion profile
behaves as
\begin{equation}
    	\sigma_\txp^2(R)
    	\approx
	\frac{8}{15\pi R}
	\left[
	1-
	\frac{15\pi^2(4-\mu)-512}{128\pi\,R}
	+\cdots
	\right],	
\end{equation}
whereas at small projected radii we obtain
\begin{equation}
    	\sigma_\txp^2(R)
    	\approx
	\frac{\Gamma^2}{6\sqrt{2}\pi^{3/2}}\,
	\frac{1-\mu}{\sqrt{R}}
	\left[
	1-\frac{3(8\pi^3-\Gamma^4)}{16\pi^2}\,R
	+\cdots
	\right]
	+
	\frac{\Gamma^4}{84\pi^2}\,\frac{\mu}{R}
	\left[
	1-\frac{3(1344\pi^4-5\Gamma^8)}{\pi^2\Gamma^4}\,R
	+\cdots
	\right].
\end{equation}
In the first panel of figure~{\ref{all.eps}} we plot the projected
velocity dispersion profile for various values of the black hole
mass. Notice that the projected velocity dispersion profile is nearly
similar to the intrinsic dispersion profile. At a given projected
radius $R$, the projected dispersion is a weighted mean of the
dispersion along the line of sight. As the density is a strongly
decreasing function of radius, stars around the tangent point
$r\approx R$ will strongly dominate this mean, such that the projected
velocity dispersion is nearly equal to the intrinsic dispersion at
this point. As a result of this similarity, the effect of a black hole
on the projected velocity dispersion is obviously similar as the
effect on the intrinsic velocity dispersion. At large projected radii,
there is obviously no effect from the black hole, whereas a black hole
changes the slope of the projected dispersion profile at small
projected radii from $-\frac{1}{4}$ to $-\frac{1}{2}$.

\section{Extension to models with an anisotropic orbital structure}

The models in the family we have described all have an isotropic
dynamical structure, and consequently a distribution function
$f(\calE)$ that depends only on the binding energy. General
anisotropic distribution function in spherically symmetric systems
have distribution functions that depend on two integrals of motion,
usually the binding energy and the modulus $L$ of the angular momentum
vector. The construction of general anisotropic dynamical models
$f(\calE,L)$ that correspond to a given spherical potential-density
pair is discussed in detail by Dejonghe~(1986). Also for anisotropic
models, the key ingredient for the calculation of the distribution
function and its moments is an augmented density. For anisotropic
models, the augmented density is an explicit function of both radius
and potential, i.e. a function $\tilde\rho(\Psi,r)$. For each
potential-density pair, infinitely many of these augmented density
functions and thus anisotropic dynamical models can be
constructed. The only requirements are that the condition
\begin{equation}
	\tilde\rho(\Psi(r),r)\equiv\rho(r)
\label{vwani}
\end{equation}
is satisfied and that the corresponding distribution function is
non-negative over the entire phase space. Unfortunately the formulae
to calculate the distribution function and its moments from the
augmented density are significantly more complicated than the
Eddington formula~(\ref{eddington}) in the isotropic case. A number of
different approaches exist, including for example an approach with
combined Laplace-Mellin transforms. There are a number of special
cases, however, for which the construction of anisotropic dynamical
models is not much more demanding than the construction of isotropic
models. This has been illustrated by Baes \& Dejonghe~(2002), who
constructed three different families of analytical dynamical models
for the selfconsistent Hernquist potential-density pair, with widely
different dynamical structure. In this section we shortly describe how
our family of isotropic dynamical models with a central black hole can
be generalized to models with a constant anisotropy, models with an
Osipkov-Merritt type distribution function and models with a
completely radial orbital structure. Further generalizations, such as
Cuddeford~(1991) models, are also possible.

\subsection{Models with a constant anisotropy}

A special family of dynamical models corresponds to models with a
augmented density $\tilde\rho(\Psi,r)$ that is a power law of $r$,
\begin{equation}
	\tilde\rho(\Psi,r)
	=
	\tilde\rho_\txA(\Psi)\,r^{-2\beta},
\label{constani}
\end{equation}
where $\beta<1$ and where the function $\tilde\rho_\txA(\Psi)$
is determined by the condition~(\ref{vwani}). It is well-known that
augmented densities of the form~(\ref{constani}) correspond to models
with a constant anisotropy (Dejonghe~1986; Binney \&
Tremaine~1987). The distribution function is a power law of the
angular momentum, and can be found through an Eddington-like formula
\begin{equation}
	f(\calE,L)
	=
	\frac{2^\beta}{(2\pi)^{3/2}}\,
	\frac{L^{-2\beta}}{\Gamma(1-\beta)\,\Gamma(\tfrac{1}{2}+\beta)}\,
	\int_0^\calE
	\frac{\txd^2\tilde\rho_\txA}{\txd\Psi^2}\,
	\frac{\txd\Psi}{(\calE-\Psi)^{1/2-\beta}}.
\label{dfca}
\end{equation}
To calculate the distribution function for the model defined by the
potential-density pair (\ref{density})-(\ref{potential}), we can
transform equation~(\ref{dfca}) to an integration with respect to $a$
in the same way as in section~{\ref{df.sec}}. We directly obtain
\begin{equation}
	\tilde\rho_\txA(a)
	=
	\frac{1}{8\pi}\,
	\frac{(a^2-1)^{4-2\beta}}{a^3}.
\end{equation}
If $\beta$ is a half-integer number, the integrand in
formula~(\ref{dfca}) is a rational function of $a$, and if $\beta$ is
an integer number, the integrand is the product of a rational function
and a square root of a quadratic term in $a$. The distribution
functions can consequently be written in terms of elementary functions
only for all integer and half-integer values of $\beta$. The radial
and tangential velocity dispersions can be found through a formula
similar to equation~(\ref{jeans})
\begin{equation}
	\sigma_r^2(r)
	=
	\frac{\sigma_\theta^2(r)}{1-\beta}	
	=
	\frac{\sigma_\phi^2(r)}{1-\beta}	
	=
	\frac{1}{\rho_\txA(r)}
	\int_r^\infty
	\frac{M(r)\,\rho_\txA(r)\,\txd r}{r^2}.
\label{cvd}
\end{equation}
These expressions can also be written in terms of elementary functions
for all integer and half-integer values of $\beta$. Similarly,
analytical expressions can be derived for all higher-order moments if
$\beta$ is an integer or half-integer number.

\subsection{Osipkov-Merritt models}

Osipkov~(1979) and Merritt~(1985) developed an inversion technique for
another special class of spherical anisotropic dynamical models,
namely models where the distribution function depends on energy and
angular momentum only through the combination $Q\equiv\calE
-L^2/2r_\txa^2$, with $r_\txa$ the so-called anisotropy radius and
under the assumption that $f(\calE,L)=0$ for $Q<0$. Such models are
characterized by an augmented density of the form
\begin{equation}
	\tilde\rho(\Psi,r)
	=
	\left(1+\frac{r^2}{r_\txa^2}\right)^{-1}
	\tilde\rho_\txQ(\Psi),
\end{equation}
where the function $\tilde\rho_\txQ(\Psi)$ is determined by the
condition~(\ref{vwani}). The Osipkov-Merritt models have an anisotropy
profile of the form $\beta(r)=r^2/(r^2+r_\txa^2)$, i.e. they are
isotropic in the centre and become completely radially anisotropic at
large radii. For the calculation of distribution functions for
Osipkov-Merritt models, one can use a formula very similar to the
Eddington formula,
\begin{equation}
	f(\calE,L)
	=
        \frac{1}{\sqrt{8}\pi^2}
        \int_0^Q 
	\frac{\txd^2\tilde\rho_\txQ}{\txd\Psi^2}\,
        \frac{\txd\Psi}{\sqrt{Q-\Psi}}.
\label{dfQ}
\end{equation}
For the model we consider, we obtain
\begin{equation}
	\tilde\rho_\txQ(a)
	=
	\tilde\rho(a)
	+
	\frac{1}{8\pi r_\txa^2}\,\frac{(a^2-1)^2}{a^3}.
\end{equation}
The distribution function of the Osipkov-Merritt models can be
calculated in a very similar way as the distribution function of the
isotropic models. We find after some algebra
\begin{multline}
	f(\calE,L)
	=
	f(Q)
	+
	\frac{\sqrt{2}}{128\pi^3r_\txa^2}
	\left[
	\frac{\calX_\txQ(Q,\mu)\,\sqrt{Q}}{(2-\mu+Q)^4\,(1+\mu Q)}
	\right.
	\\
	\left.
	+\frac{3\,(1-\mu)\,(19-24\mu+16\mu^2-16Q+23\mu Q-4Q^2)}
	{(2-\mu+Q)^{9/2}}\,
	\arctanh\left(\frac{\sqrt{(2-\mu+Q)\,Q}}{1+Q}\right)\right],
\end{multline}
where $f(Q)$ represents the isotropic distribution
function~(\ref{fBH}) and $\calX_\txQ(Q,\mu)$ is a polynomial with
integer coefficients
\begin{equation}
	\calX_\txQ(Q,\mu)
	=
	(199-383\mu+264\mu^2-80\mu^3+16\mu^4)
	+
	(67+58\mu-189\mu^2+96\mu^3)\,Q
	+(14+7\mu-5\mu^2)\,Q^2
	+2(1-\mu)\,Q^3.
\end{equation}
The velocity dispersions for the Osipkov-Merritt models can be found
through the formula
\begin{equation}
	\sigma_r^2(r)
	=
	\frac{\sigma_\theta^2(r)}{1-\beta(r)}	
	=
	\frac{\sigma_\phi^2(r)}{1-\beta(r)}	
	=
	\frac{1}{\rho_\txQ(r)}
	\int_r^\infty
	\frac{M(r)\,\rho_\txQ(r)\,\txd r}{r^2}.
\end{equation}
This integral can easily be evaluated analytically, and also the
higher-order moments of the distribution function can be expressed
completely in terms of elementary functions for all values of
$r_\txa$ and $\mu$.

\subsection{Models with a completely radial orbital structure}

It deserves some special attention that the distribution function of
both the models with constant anisotropy and the Osipkov-Merritt
models corresponding to our potential-density pair
(\ref{density})-(\ref{potential}) are positive for all values of
$\beta$ and $r_\txa$ and for all values of the relative black hole
mass $\mu$. This is quite unusual. In the limit $r_\txa\rightarrow0$
or $\beta\rightarrow1$ for the Osipkov-Merritt and constant anisotropy
models respectively, the entire galaxy is populated with only radial
orbits. Not all density profiles can sustain such a model: Richstone
\& Tremaine~(1984) demonstrated that the density must diverge at least
as $r^{-2}$ at small radii in order to be able to sustain a purely
radial orbital structure. Models with constant density cores or
density cusps shallower than $r^{-2}$ will hence have a minimal
anisotropy radius $r_{\text{a,min}}$ below which Osipkov-Merritt
models are negative at some point in phase space, and a maximum
anisotropy value $\beta_{\text{max}}$ above which the constant
anisotropy models become negative at some point in phase space. For
example, selfconsistent Plummer models have $\beta_{\text{max}}=0$ and
$r_{\text{a,min}}=\tfrac{3}{4}$ (Merritt~1985), and selfconsistent
Hernquist models have $\beta_{\text{max}}=\tfrac{1}{2}$ and
$r_{\text{a,min}}\approx0.202$ (Baes \& Dejonghe~2002). The central
slope of the density~(\ref{density}) for the models we consider in
this paper is steep enough such that even models with a purely radial
orbital structure are positive over the entire phase space.

For models with a completely radial orbital structure, the augmented
density can be written as
\begin{equation}
	\tilde\rho(\Psi,r)
	=
	\frac{\tilde\rho_\txR(\Psi)}{r^2},
\end{equation}
where the function $\tilde\rho_\txR(\Psi)$ is determined by the
condition~(\ref{vwani}). The distribution function is of course a
degenerate function of angular momentum (only orbits with $L=0$ are
populated), and can be calculated via the Eddington-like formula
\begin{equation}
	f(\calE,L)
	=
	\frac{\delta(L^2)}{\sqrt{2}\pi^2}
	\int_0^\calE
	\frac{\txd\tilde\rho_\txR}{\txd\Psi}\,
	\frac{\txd\Psi}{\sqrt{\calE-\Psi}}.
\end{equation}
Once again, a closed expression can be obtained for the family of
potential-density pairs we have discussed in this paper by
transforming the integral to an integration with respect to the
variable $a$. One obtains after some algebra
\begin{multline}
    	f(\calE,L)
    	=
    	\frac{\sqrt{2}\,\delta(L^2)}{32\pi^3}
	\left[
	-
	\frac{(17-29\mu+14\mu^2-3\calE+3\mu\calE-2\calE^2)\,\sqrt{\calE}}
    	{(2-\mu+\calE)^3}
	+
    	\frac{2}{\sqrt{\mu}}
    	\arctan\sqrt{\mu\calE}
    	\right.
	\\
	+
    	\left.
	\frac{(1-\mu)\,(1-4\mu-2\mu^2-16\calE-17\mu\calE+4\calE^2)}
    	{(2-\mu+\calE)^{7/2}}
	\arctanh\left(\frac{\sqrt{(2-\mu+\calE)\calE}}{1+\calE}\right)
    	\right],
\end{multline}
This distribution function is everywhere positive, for all values of
the black hole mass $\mu$. 
The tangential velocity dispersions for these models are of course
identically zero, whereas the radial velocity dispersion can be found
through equation~(\ref{cvd}),
\begin{equation}
	\sigma_r^2(r)
	=
	(1-\mu)\sqrt{\frac{1+r}{r}}
	\left(1+2r-4r(1+r)\ln\sqrt{\frac{1+r}{r}}\right)
	+
	\frac{2\mu}{3}\left(\frac{1+r}{r}\right)
	\left(1-4r-8r^2+8r^{3/2}\sqrt{1+r}\right).
\end{equation}

\section{Conclusion}

We have described a one-parameter family of spherical models for
elliptical galaxies and bulges consisting of a stellar component and a
central black hole. All models in this family share the same stellar
density profile, which has a steep cusp with a slope of
$-\tfrac{5}{2}$ in the centre. The gravitational potential for the
models is a linear combination of the potential generated
selfconsistently by the stars and the potential of a central
supermassive black hole. The parameter $\mu$, representing the mass of
the black hole relative to the total mass of the galaxy, can assume
any value between 0 and 1. The family therefore contains models
ranging from a selfconsistent model without black hole to a model
where the potential is entirely dominated by the black hole. We have
done an extensive study of the internal and projected kinematics for
this family of models. With the assumption of isotropy, we have
calculated the intrinsic velocity dispersions, the distribution
function, the differential energy distribution, the moments of the
distribution function, the surface brightness and the projected
velocity dispersions. All of these quantities have been expressed
completely analytically, for all values of $\mu$.

We have also described some extensions of the models to anisotropic
orbital structures. In particular, we have considered models with a
constant anisotropy, models with distribution functions of the
Osipkov-Merritt type and models with a completely radial orbital
structure. Also for these families of models, the distribution
function and its moments can be calculated completely analytically for
all values of the central black hole mass $\mu$.

We are well aware that the stellar component we have considered does
not completely represent the detailed structure of the centre of
galaxies. Firstly, it has a fairly steep density cusp, which is at the
edge of the observed range in real elliptical galaxies (Lauer et
al.~1995; Gebhardt et al.~1996). With such a steep cusp, much of the
stellar mass is very centrally concentrated, resulting in an
infinitely deep stellar potential well. The addition of a black hole
potential to this stellar potential does not drastically change the
global mass distribution, such that the effects of a black hole on the
kinematical properties are probably quite conservative. For example,
the distribution function is substantially, but not very drastically
affected by the presence of a black hole, whereas this effect is much
stronger for $\gamma$-models with a less steep density cusp (see
figure~6 in Tremaine et al.~1994). The reason why we focused on the
density profile~(\ref{density}), and not on one of the other members
of the family of $\gamma$-models with a shallower central density cusp
for example, is that the $\gamma=\tfrac{5}{2}$ model is unique in the
way that it allows a relatively simple expression of the augmented
density in the presence of a central black hole. It is the only member
of the family of $\gamma$-models where the distribution function, the
differential energy distribution and the moments can completely be
expressed in terms of elementary functions. We are currently
undertaking a more general study of the properties of the
$\gamma$-models with a central black hole, using both analytical and
numerical means (Baes, Buyle \& Dejonghe~2004).

Secondly, the models presented here are spherically symmetric and
isotropic, whereas few elliptical galaxies are thought to be perfectly
spherical. Observational studies suggest that a substantial fraction
of elliptical galaxies are at least moderately triaxial (Franx, van
Gorkum \& de Zeeuw~1991; Tremblay \& Merritt~1995; Bak \&
Statler~2000). The central regions of galaxies with a central black
hole are generally believed to be roughly axisymmetric,
however. Indeed, supermassive black holes drive the shape of galaxies
in a triaxial haloes toward axisymmetry by stochastic diffusion,
either globally (Gerhard~\& Binney~1985; Norman, May \& van
Albada~1985; Merritt \& Quinlan~1997; Wachlin \& Ferratt-Mello~1998;
Valluri \& Merritt~1998), or at least the central regions
(Holley-Bockelmann et al.~2001, 2002). Still, axisymmetric systems
have a much larger freedom in orbital structure than spherical
systems. Using detailed axisymmetric dynamical modelling, Gebhardt et
al.~(2003) found that the central regions of the massive early-type
galaxies generally have a significant tangential anisotropy, whereas
less massive galaxies have a range of anisotropies.

In spite of these critical notes, the family of dynamical models
presented in this paper has the huge advantage over more complicated
numerical models that all kinematical properties can be calculated
completely in terms of elementary functions, for any value of the
parameter $\mu$. This is not only the case for rather simple
kinematical properties such as the velocity dispersions, but also for
more complicated kinematical properties which depend in a strongly
non-linear way on the potential. In particular, the distribution
function and the differential energy distribution can be written in a
compact form and in terms of elementary functions only. As a result,
this family of models is useful for a large set of applications. In
particular, it can be used to easily generate the initial conditions
for N-body or hydrodynamical simulations, which are needed to
investigate how black holes interact with the stellar, gaseous and
dark matter components of their host galaxies.

\section*{Acknowledgement}

Part of this work was done in Cardiff, where M.B. was a visiting
postdoctoral fellow. He gratefully acknowledges the hospitality of
Cardiff University and the financial support of the Fund for
Scientific Research Flanders. The authors thank the anonymous referee
for useful suggestions that enhanced the contribution of the paper.

\appendix

\section{Evaluation of the incomplete elliptic integrals}

The expressions~(\ref{Sigmahere}), (\ref{SR}) and (\ref{sigp}) for the
surface brightness, the cumulative surface brightness and the
projected velocity dispersion contain incomplete elliptic integrals of
the first and second kind. When the modulus of these functions (the
second parameter) is less than unity, these functions are easily
evaluated numerically, and simple and efficient routines are widely
available (e.g.\ Press et al.~2001). Sophisticated closed-box
mathematical packages such as Maple or Mathematica easily calculate
incomplete elliptic integrals for all (complex) values of the modulus,
but most basic implementations break down when the modulus is greater
than unity. For the present case, this happens when $R<1$. A
well-behaved solution for $R<1$ can be obtained with the transformtion
formulae for incomplete elliptic integrals (e.g.\ Abramowitz \&
Stegun~1972). The relevant formulae to transform the expressions for
the surface brightness profile etc.\ for $R<1$ are
\begin{gather}
    	\mathbb{F}\left(\frac{\pi}{4},\sqrt{\frac{2}{R+1}}\right)
    	=
    	\sqrt{\frac{R+1}{2}}\,
    	\mathbb{F}\left(\arctan\sqrt{\frac{1}{R}},
    	\sqrt{\dfrac{R+1}{2}}\right),
    	\\
    	\mathbb{E}\left(\frac{\pi}{4},\sqrt{\frac{2}{R+1}}\right)
    	=
    	\sqrt{\frac{2}{R+1}}\,
    	\mathbb{E}\left(\arctan\sqrt{\frac{1}{R}},
    	\sqrt{\dfrac{R+1}{2}}\right)
    	+
    	\frac{R-1}{\sqrt{2(R+1)}}\,
    	\mathbb{F}\left(\arctan\sqrt{\frac{1}{R}},
    	\sqrt{\dfrac{R+1}{2}}\right).
\end{gather}
Around $R=1$ either of these forms are hard to evaluate numerically,
because the modulus of the elliptic integrals then approaches
unity. Instead, it is better to use the Taylor expansion
\begin{gather}
    	\mathbb{F}\left(\frac{\pi}{4},\sqrt{\frac{2}{R+1}}\right)
    	\approx
	\frac{\ell}{\sqrt{2}}
	+
	\frac{\ell-2}{8\sqrt{2}}\,(R-1)
	-
	\frac{7\ell-26}{256\sqrt{2}}\,(R-1)^2
	+
	\cdots,
	\\
    	\mathbb{E}\left(\frac{\pi}{4},\sqrt{\frac{2}{R+1}}\right)
    	\approx
	\frac{1}{\sqrt{2}}
	+
	\frac{\ell-1}{4\sqrt{2}}\,(R-1)
	-
	\frac{5\ell-4}{64\sqrt{2}}\,(R-1)^2
	+
	\cdots,
\end{gather}
where
\begin{equation}
	\ell 
	=
	\frac{1}{\sqrt{2}}\arctanh\left(\frac{1}{\sqrt{2}}\right)
	=
	\frac{1}{2\sqrt{2}}\ln\left(\frac{\sqrt{2}-1}{\sqrt{2}+1}\right)
	\approx
	0.62323.
\end{equation}
For example, for the surface brightness profile, we obtain
\begin{equation}
	I(R)
	\approx
	\frac{1}{\pi}
	\left[
	\frac{10-7\ell}{16\pi}
	-
	\frac{3\,(62-37\ell)}{256\pi}\,(R-1)
	+
	\frac{3454-1797\ell}{4096\pi}\,(R-1)^2
	-
	\cdots
	\right].
\end{equation}

\end{document}